# A Security Analysis Of Browser Extensions

## Authors

- Abhay Rana (nemo@sdslabs.co.in)
- Rushil Nagda (rushil9292@gmail.com)

## Abstract

Browser Extensions (often called plugins or addons) are small pieces of code that let developers add additional functionality to the browser. However, with extensions comes a security price: the user must trust the developer. We look at ways in which this trust can be broken and malicious extensions installed. We also look at silent installations of plugins in various browsers and work on ways to make silent installations possible in browsers that work against it.

We compare the browser extension mechanism among various browsers, and try to create a set of rules to maintain the principle of least privileges in the browser. We track various plugins and determine whether the least privileges required match with the privileges asked for.

We also work on a survey of extensions (for various browsers) and determine the nature of attacks possible. For eg, if a developer account gets hacked, updating of a normal extension with a malicious one is possible. We look at privilege abuse and survey extensions that ask for more privileges than they use.

We finally provide a solution and allow a person to check the authenticity of the extension even before they download it.

## 1. Introduction

Extensions have grown to be extremely popular among users, with over 33% of Google chrome users having at least one extension installed.[3] Extensions(also known as plugins), are tiny pluggable pieces of code that allow a user to modify the browser's behaviour, on certain (possibly all) websites. Most extensions are written in javascript, although there are some exceptions. While extension usage is widespread on the desktop, it is still in it's native stage on mobile browsers. Only a few mobile browsers, such as Mozilla's Firefox for Android and Dolphin support third party extensions.

In this paper, we limit ourselves to Google Chrome & Firefox (both Desktop versions), the world's leading browsers. We couldn't work on other browsers due to the time-constraint imposed on our study.

Most browsers have some form of a security model on their extension platform. In this paper, we analyse the different vulnerabilities an extension could cause and/or succumb to. We examine the effectiveness of the security model based on our results.

## 2. Security Overview

Google Chrome uses three security concepts for keeping extensions secured:

- *Isolated Words*. An extension's content scripts cannot access the direct DOM (Document Object Model) of the current running page, but access a copy of it. The javascript execution of content-scripts is kept completely separate from the execution of the page's actual javascript code, if any. This leads to a large number of attack vectors getting nullified.
- *Privilege Separation*. Google Chrome extensions are run in two different privilege modes. One is the content-scripts, and the other being core-extension scripts. Core extension scripts have access to the chrome native APIs, while content-scripts do not. Content-scripts however, can access those APIs by using a message passing interface to *talk to* the core-extension scripts. This leads to strengthening the security model by making a vulnerability in the messaging interface necessary before the browser can be attacked.
- *Permissions Model*. Chrome extensions come with a privilege model, where extensions are required to pre-declare their needed privileges, and are limited to those by the browser. If a vulnerability is found in the core-extension, the attacker will have still be limited by the privileges the extension has attained.

An overview of Chrome extensions is shown below:

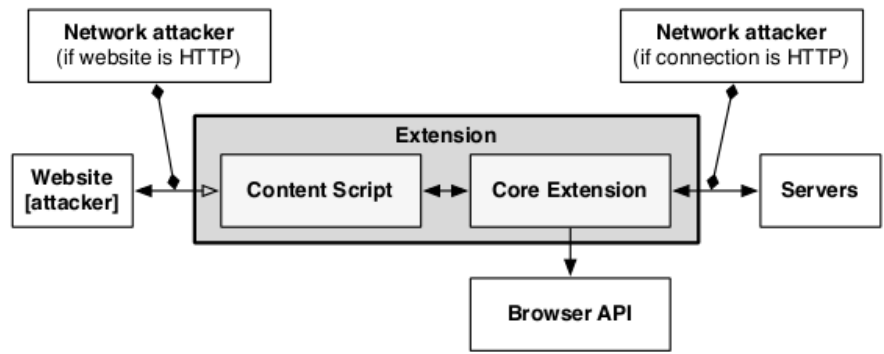

Figure 1: The architecture of a Google Chrome extension.

## Threats

There are two possible attacks on extensions in a browser:

1. Malicious Extensions: An attacker could install a malicious extension in the browser that could, theoretically, cause a lot of damage. A few malicious extensions have been found in the wild before, and they were used to steal passwords for Banking and other High Security sites. These were promptly blacklisted by the Browsers themselves, but the threat still remains.
We look at malicious extension installation mechanisms, via which an executable can install an extension silently. Both Google Chrome and Firefox provide a method to install extensions silently. However, both of these show a confirmation dialog to the user before completing the installation.
2. Extension Vulnerabilities: The extension itself may suffer from one or more vulnerabilities, due the insecure coding practices, or because simply, the developer does not know any better. There are more than 10,000 extensions on the Chrome Web Store [4], and some of them are bound to suffer from vulnerabilities. A previous study[1] has shown that even the most popular extensions, including those developed by Google, suffer from many vulnerabilities.
These extension vulnerabilities can be exploited by malicious websites. This may include, improper handling of user input, or unsanitized input being used in the extension, and metadata attacks on the extension.

We do not do a study on the Firefox Addon Store because of the fact that Firefox does not have a permissions management system for extensions. As a result of this, all extensions have full read-write privileges on the user's home directory. This makes

Firefox extensions a lot more powerful, but also more vulnerable at the same time, since an extension vulnerability could lead to malicious code execution on the machine.

## 3. Methodology

We perform the following:

1. Create a method to bypass the "prompts" that a browser give for silent extension installations. We perform this for both Firefox and Chrome.
2. Analyse the code of the top 10,000 extensions on the Google Chrome Web Store. We analyse the code statically, and work on some attack vectors and privilege abuse.
3. Create a method to view the results of our Step 2 above, before installing a new extension. This would allow a user to analyse the extension better before deciding to install it. Our study shows that several extensions, ask for more privileges than they use.

## 4. Statistics

We downloaded a total of 10,047 extensions from the Chrome Web Store. They were downloaded in order of popularity, so these formed the most used extensions list on Chrome.

**Content-Security Policy**

Chrome version 18 brought a Content-Security Policy feature to Chrome Extensions. This prevents most of the vulnerabilities we discuss, by enforcing a stricter policy. We found that 4079 extensions had upgraded to manifest version 2, thus enforcing CSP upon them. These extensions were still, however, available for privilege abuse and some vulnerabilities. These extensions were, in general, found to be more secure than the others.
Out of our 10,047 downloaded extensions, we were able to scan only 9558 extensions, due to errors such as improper manifest file or missing files.

**Privilege Abuse**

We scanned our set of extensions for the permissions they asked of the user. This was accomplished by parsing through their 'manifest.json' files. From the google chrome developer page, '*Every extension, installable web app, and theme has a JSON-formatted manifest file, named manifest.l, or other files missing.json, that provides important information.*'

We then matched these with the permissions the extensions actually made use of. The chrome security model only allows scripts listed as 'background-scripts' to access core API functions. Parsing through the source codes of all the files listed as background-scripts, we were able to determine the permissions that these extensions really needed.

We found that almost half the extensions in our set asked for more permissions than they used. Most extensions asked for just one or two extra permissions, and this can be attributed to shoddy developing or negligence. 192 applications asked for 4 or more permissions than they used and these represent serious security issues.

We are not, in our statistics, including extra permissions that cannot be abused. The 'notifications' permission for example, allows the extension to use the HTML5 notification api and cannot be abused. Over 300 applications that asked for this privilege without using it have not been listed in our statistics. While it is a bad practice to ask for these permissions, they do not pose any sort of a security threat.

The data we managed to gather:

| Number of extra privileges sought | Number of violating extensions |
| --- | --- |
| 1 | 3237 |
| 2 | 923 |
| 3 | 250 |
| 4 | 92 |
| 5 | 52 |
| 6 | 19 |
| 7 | 5 |
| 8 | 6 |
| 9 | 9 |
| 10 | 0 |
| 11 | 0 |
| 12 | 1 |
| 13 | 2 |
| 14 | 3 |
| 15 | 1 |
| 16 | 2 |

Here is a plot of number of violating extensions vs number of extra permissions they ask for. The y-axis is plotted on a logarithmic scale of base 10.

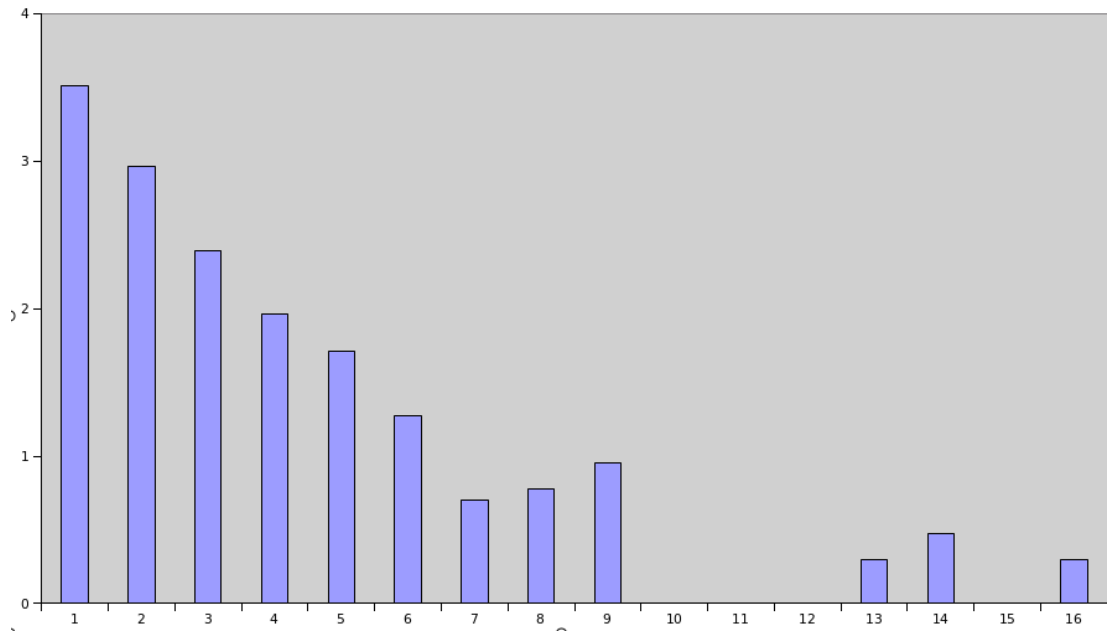

Many of the violating extensions ask for extremely sensitive permissions and can be abused easily, by both, an ill-intentioned developer or a hacker who has gained access to a developer's account. For example, some of these extensions ask for access to a user's browser cookies. With access to these, an attacker could potentially gain access to a user's online accounts.

## Network Attacks

We found that some extensions were vulnerable to HTTP-network attacks, as they were loading scripts from an HTTP domain. We were not able to count *every* HTTP request made, due to technical reasons, but found that at least 146 extensions made a request to an HTTP address. Since this only includes direct `<script>` requests, this number would have been much higher, if we had also included XHR requests (which many extensions used).

Counting the XHR requests would have been a herculean task to do manually, and we were not able to find a way to perform it using code analysis. This was partially because of the many number of ways a network call could have been made in Javascript, from including an IMG tag to performing XHR requests using JQuery, MooTools, or any other JS library. We contemplated checking all URL strings as *"requested"* urls, but decided against it as it would have resulted in a large number of false positives.

Network attacks over HTTP includes a *Man in the Middle* (MitM) attack, where a malicious machine may change the javascript en-route to the user. This allows remote code execution, possible on the core-extension part, which would mean all privileges by the extension can be abused. For instance, a malicious MITM attack might copy over all of user's cookies and send them off to the attacker; this would result in a *Holy Grail* attack with all of user's accounts getting compromised at once.

# Working Model

### Silent Installation

Firefox and Chrome have methods for Silent Extension installations. These involve registry editing under windows, as the norm. The browser, upon restart, detects the registry entry and shows a warning popup to the user about the installation. The installation is only allowed, if the user permits is on this warning window.

We create a mechanism to install an extension which does not raise this warning screen. We do this via different methods on each browser:

1. *Firefox*: Firefox keeps a list of installed extensions in a sqlite file called *extensions.sqlite* inside the user's profile directory. We directly write a new entry to this database "faking the installation procudure". This method was based off earlier work done on this by Julien Sobrier[2]. This allows us to easily install an extension, without triggering the browser warning popup window.
2. *Google Chrome*: Chrome keeps the extension installation preferences inside the*Preferences* file in the user data directory. This file is created in JSON format. We install an extension by creating a fake entry in this JSON file. Upon restart, the extension is shown as installed in the browser.

Our work on Silent Installation Mechanism proves that the current implementations are not strong enough and can be bypassed easily. Even though, it would forever be a game of cat-and-mouse with malicious extension developers trying to keep up with the latest anti-silent measure introduced by the browser.

**Extension Checker**

We create a method to allow users use our research by setting up an online database with results of our analysis. A user can easily lookup online the extent of permission abuse or HTTP traffic that the extension makes. We plan to put up the website online soon.

The website works by comparing the permissions asked during installation by reading the `manifest.json` file and the permissions used, which we've analysed once and put in our database. We also show the network HTTP usage of the extension by checking up our database (which holds the values collected during the scan).

## Conclusion

We successfully develop scripts for silent installations of extensions in both Firefox and Google Chrome, showing that the current mechanisms in force to prevent this are not effective.

Close to 50% of the extensions we scrutinised have some form of privilege abuse. While it is difficult to put an exact number on how many of these are serious potential threats, we do conclude there are a large number of extensions susceptible to attacks via privilege abuse.

We also realize that the Content-Security-Policy is quite effective at making extensions more secure, and that the Chrome Web Store will be much safer as it is enforced more strictly over time.

# Appendix

All source code, including list of extensions downloaded for the study, is available on request at nemo@sdslabs.co.in